# Promoting Game Accessibility: Experiencing an Induction on Inclusive Design Practice at the Global Games Jam

Michael James Scott
Information Systems, Computing & Mathematics
Brunel University
Uxbridge, Middlesex, UB8 3PH
United Kingdom
michael.scott@brunel.ac.uk

Gheorghita Ghinea
Information Systems, Computing & Mathematics
Brunel University
Uxbridge, Middlesex, UB8 3PH
United Kingdom
george.ghinea@brunel.ac.uk

## ABSTRACT
The Global Games Jam (GGJ) attracts many people who are passionate about games development, coming from a range of educational backgrounds. Therefore, the event can be experienced by novices and student developers as an opportunity for learning. This provides an opening to promote themes and ideas that could help form future thinking about games design, emerging as a form of induction on key design issues for new practitioners. Such an approach aims to raise awareness about issues which learners could help develop and take with them into industry. However, the experience itself affords a deep experiential rhetoric and dialogue with experts that could be an effective pedagogical tool for issues seldom addressed deeply in formal educational settings. This paper describes an account by one such individual, being introduced to game accessibility through participation in the GGJ. As such, it is not intended as a rigorous empirical analysis, but rather a perspective on one way a game jam can be experienced, inviting further research on the topic.

## Categories and Subject Descriptors
K.3.2 [**Computers and Education**]: Computer and Information Science Education – *computer science education, information science education.*

## General Terms: Human Factors

## 1. INTRODUCTION
As a participant in the Global Games Jam, the author found that the event attracted a fair proportion of novices interested in games development. Furthermore, a number of sites are based in educational institutions where the resident students were encouraged to attend. It would, therefore, be prudent to explore the benefits of this hackathon-like event from an educational perspective, focusing on how it has been used to connect learners with new learning opportunities.

These types of event are not only an opportunity to test and develop skills, but also an opportunity to learn from other developers. The emphasis being on peer-supported practical activity and collaboration, representing a refreshing change from the traditional pedagogies and curricula of formal institutions. Furthermore, as a student, the notion of learning from industry practitioners is rather attractive because of the perceived legitimacy and relevance. Even if it turns out that they do not work directly together, the discourse and critique that is often encouraged between teams is appreciated.

One way in which the Global Games Jam has differentiated itself from other hackathons, has been in the diversity and flexibility it welcomes. The event allows for site-specific local constraints, which some locations have embraced to explore key design issues, pertinent to the future of games. Working alongside practitioners and advocates, optional challenges are presented at some venues, forming a type of induction on design practice for those new to the field. The term 'induction' being used deliberately because, as this paper argues, the unique culture at the event can achieve more than just promotion.

In 2012, arriving at the at the SAE Institute in the United Kingdom (UK), the author experienced such an induction, being introduced to game accessibility in the form of the Games Accessibility Challenge. The objective: to produce a game design that addresses a range of accessibility pitfalls.

## 2. PROMOTING ACCESSIBILE DESIGN
From a student perspective, the role of the challenge initially seemed to just be to promote awareness of the issue of game accessibility. The term is sometimes used by industry practitioners to refer to the availability of a title, or the ease of play for a lay individual. Often, however, this is not what advocates mean. To clarify, it is defined here as the application of inclusive design practices in order to remove barriers to playing games that can affect individuals with impairments.

The accessibility challenge was introduced by the organizers after the theme of the Game Jam was revealed. It was emphasized that it was optional, but a brief argument was put forward as to why accessibility was important for game designers. Intrigued by these claims, participants arranged themselves into teams with many discussing the challenge. This prompted one participant to briefly consult an open-access literature review in the area [23], which yielded some interesting findings relating to how the issue affects many different players.

According to a 2007 survey conducted by the NPD Group, more than 100 million consoles are present in the United States (US) [12] and the ESA also estimated that 63% of the U.S population plays video games, with over 50% of players doing so on a weekly basis [8]. Despite this apparent popularity, however, a large group of people find themselves excluded from this cultural phenomenon because of an impairment [1, 2]. The term "digital outcasts" [21, 22] has emerged to describe these individuals, who seem to be left behind by the rapid evolution of, often somewhat inaccessible, technological innovations.

Contrary to popular belief, the number of people affected is not trivial. Based on the 2002 US Census [18], Yuan et al [23] estimated that the ability of at least 32 million people to play games could be affected. This is equivalent to ~11% of the US population, which can be further detailed as about 9% who are able to play with a reduced gaming experience and approximately 2% who want to play games but are completely unable to. Furthermore, Smith [15:122] notes that many of those affected do play as, according to a 2008 survey of PopCap users by the Information Solutions Group [13], "over a fifth of game players self-identified as having some form of a disability". Therefore, game accessibility is an important issue for industry, from a commercial as well as a social perspective, as digital outcasts are a market that is currently underserved by mainstream offerings.

It should also be noted that games are increasingly being applied in contexts beyond entertainment. For example, educators have explored the use of games in the classroom environment [16]. Historically, however, there have been legal obligations in place to ensure accessibility. For example, Section 508 of the US Rehabilitation Act in 1998 [19] stated that all educational institutions which depend upon federal funding must make their information technologies accessible. Similar legislation also exists in other nations, such as the 1998 Disabilities Act in the United Kingdom (UK) [3]. Thus, inclusive design practices are an important consideration for aspiring game developers.

While a range of innovations in accessible design have been developed [15], Bierre et al [2:8] note that the "return on investment is crucial for any game development, [so] it goes without saying that the efforts of game accessibility must have a realistic financial grounding, otherwise they risk not [being] implemented". In addition, the notion that developers "can't control what [they] can't measure" [5] is a well-known sentiment in the computing discipline which is equally relevant in computer games development. So how could developers in industry appropriately assure game accessibility for a wide range of impairments in a cost-effective manner?

Nielsen [11:252] claims that "a good strategy for improving usability in most industrial situations is to study those usability methods which are likely to see practical use". Therefore, as the organizers of the accessibility challenge seemed to argue, minor pitfalls that affect a large number of people can be easily avoided with simple checklists and testing protocols. Their proposal had remarkable similarity to the heuristic approach to the evaluation of user interfaces, where developers and experts identify whether well-established guidelines are implemented [11]. Furthermore, this approach and can often catch a large number of problems with a small number of raters [10]. Therefore, it could be a feasible technique that some producers would consider.

Thus, at the start of the challenge, all participants were handed a sheet containing such guidelines and the told the aim was to violate as few of these guidelines as possible to score the most points in each category. The categories seemed to be based on the World Health Organization's (WHO) International Classification of Impairments, Disabilities and Handicaps [4] which includes: visual impairment; hearing impairment; motor impairment; and cognitive impairment. An additional category on awareness and documentation was also included. There were 14 guidelines in total, with examples including: "colour-blind friendly"; " reinforce important audio information with visual effects or text"; "remappable controls"; "game can be started without navigating multiple levels of menus"; and "list accessibility features and game requirements" [20].

## 3. GAME MAKING AS RHETORIC

Guided by the principles offered by the accessibility guidelines, and the mentorship of a subject specialist attending the event, the development of *Mini Free Runner* [9] came together as a means of adapting a traditionally time-based game to have no time requirements. It was inspired by the notion that the open-source indie title *CANABALT* [14] could become more accessible with an alternative mechanic to support players with motor impairments.

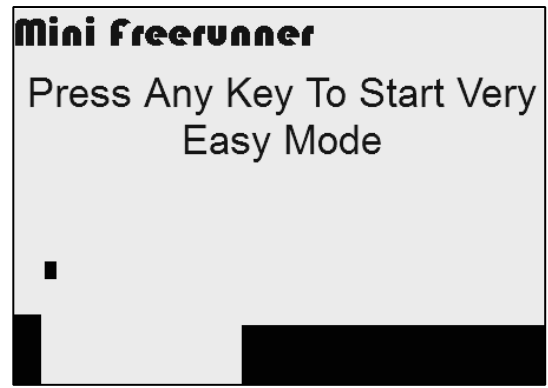

**Figure 1. Many Accessibility Options Were Added within the Time Limit, such as a Single-Switch Scanning Menu**

The aim of the game is to navigate an avatar across a series of platforms, avoiding pitfalls and obstacles by jumping as the game environment scrolls to the right. Rather than timing presses to jump at the right moment, as is required in the original, the new mechanic uses different types of jump and an intensity input to fulfill the same goal. Furthermore, as the avatar moves across the platform, it slows and then stops, before falling off the edge, providing sufficient opportunity for input. Despite this change, the gameplay remained somewhat compelling and might interest those unable to play the original game.

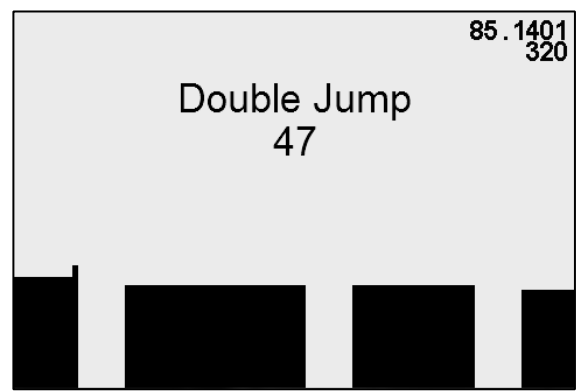

**Figure 2. Time-Sensitivity is Removed as the Game Slows and Pauses as the Avatar Approaches a Platform Edge**

While some of the more ambitious features were not implemented within the weekend prototyping period, many accessibility features did make it into the game. These included: remapping controls; text-to-speech narration; one-button mode; alternative controller support; simple start mode; practice mode and subtitled tutorial; alternatives to sound; alternatives to colour indicators; volume adjustments; noise input; and high visibility graphics. Additionally, a scanning single switch compatible menu system was partially implemented but unfinished due to the wide range of options and settings available.

Implementing all of these features during the event served as a powerful experiential rhetoric on how accessibility can be addressed within a short timeframe and with little resources when some forethought is applied. Furthermore, reviewing the progress, ideas and experiments addressed by the other teams at the venue was a source of inspiration, with the dialogue that is encouraged between participants and experts being an invaluable source for stimulating learning as well as debate about how accessibility can be addressed.

## 4. IMPACT OF THE INDUCTION

From this experience, there seem to be six interrelated areas of significant impact for students and novice game developers involved in such challenges. These being:

- becoming aware of key issues in games;
- developing key design and problem solving skills;
- appreciating pertinent issues in the industry;
- being engaged in discussion, experimentation and the proliferation of new ideas;
- observing the dissemination of new ideas;
- and broadening professional engagement with appropriate networking with participants and members of special interest groups.

Promoting key design issues is important because they can help new practitioners develop their own perspectives and take their insights with them into industry once they graduate. Furthermore, many such issues are seldom covered during formal university education; except perhaps, in the case of specialist computing courses. Even then, however, it can sometimes be the case that such topics are poorly addressed. Typically, only consisting of a small collection of lecture notes, links to further online reading materials, or a peripheral topic of discussion that emerges from practical activities.

Having learned more about the significance and complexity of an issue like accessibility, it sparks a sense of curiosity. However, traditional presentations do not often help people to enhance their design and problem solving skills. This often requires further reading and practical experience. However, during an event like the Global Games Jam, knowledge acquisition is catalyzed by advice from an expert mentor while directly working on a project provides the opportunity to develop relevant knowledge.

Furthermore, the practical undertaking can be a powerful rhetorical experience for a novice developer. As Hamilton [in 17] notes, "fully functioning and accessible games being produced in the space of 48 hours is a really powerful demonstration that accessibility doesn't have to be expensive or difficult". Thus, students can come to appreciate the challenges presented by key issues and may adopt a more positive attitude towards such challenges when they encounter them in the future. For example, after some experience using similar guidelines, participants may be more inclined to consider applying those available on web resources like Game Accessibility Guidelines [7].

The lusory context of these development activities, with their themes, diversifiers and time limits, also seems to breed creativity. This is not impaired by the the addition of extra constraints on project requirements to help that focus the scope projects on specific challenges in games design. As issues raised and discussed by teams, it fuels the proliferation of new ideas. Often, being interesting, unusual and experimental. For example, the concept of making a traditionally time-sensitive game without any twitch elements may not have been conceived outside this type of environment. Especially in a formal educational setting, where there can be a perception that risky ideas are likely to result in an unsuccessful project and therefore a poor grade. Thus, the game jam can offer a safe space for experimentation, encouraging engagement with such endeavors.

These ideas were subsequently discussed at the event during development, as discourse between teams was encouraged, and in the closing presentations. For some ideas, these discussions continued after the event between peers and mentors. Moreover, some projects and ideas receive attention on blogs and special interest websites [e.g. 6, 15], or are shared between members of special interest groups through mailing lists. Consequently, encouraging dialogue with a wider audience on how key issues could be addressed through augmenting design practice.

As a student, *being part of the experience*, seems to be very engaging and helps to develop awareness of professional issues that could be encountered in the future. The game jam can also become the beginning of a new interest. After receiving feedback from the judges of the Games Accessibility Challenge, the team was made aware of relevant professional and web communities on the topic; many, of which, the author now follows. This may be of particular interest to special interest groups who may be interested in student outreach; inspire those considering options for their final year project or masters dissertation to do something in a related area.

As an additional note, the size and awareness of the Global Games Jam are likely to be advantages that appeal to advocates of key issues in the games industry. This is because, as a student, it can be difficult to travel to off-campus bring-your-own-computer events so it often requires intrigue to motivate attendance. As such, a dedicated accessibility hackathon organized by a particular special interest group may not attract the same size and breadth of participants. Therefore, the diversity of the audience attracted to this globally recognized event potentially provides an appealing opportunity to conduct this type of outreach activity.

The overall experience would seem to be positive for both participants and hosts. However, these observations spark a range of questions for further research. For example, how effective are such inductions? Did other participants find the experience engaging? How many event attendees actually participated in the challenge? Were students the ones most likely to participate? Is the approach scalable to different venues? Are other advocacies interested and piloted at other venues? Most importantly, if such inductions are effective, how they can be effectively facilitated without undue interference to the core aims and success of the Global Games Jam?

## 5. CONCLUSION

The Global Games Jam is a growing cultural element of the professional development of many novices who are passionate about games development. Therefore, aspects of the event could be developed and leveraged to address issues pertinent to the games industry and a society that enjoys playing games. It is therefore argued that optional challenges organized by hosts, such as the Games Accessibility Challenge, could lead to learning, attitude change, experimentation, the proliferation and dissemination of novel ideas, as well as the broadening of special interest networks.

However, this perspective is based on a subjective reflection of a personal learning experience at the Global Games Jam. While it should not be considered a rigorous scientific account, the issues discussed raise a range of questions about the effectiveness and scalability of such endeavors. Further investigation using surveys, ethnography and other empirical methods such as content analysis are proposed, for an event in the future, in order to: document the role that game jams can have in raising awareness about different issues; explore predominant attitudes of participants towards such issues; and determine the effectiveness of issue induction with greater rigor.

## 6. ACKNOWLEDGEMENTS

The *Mini Free Runner* game [9] was designed and coded during the 48 hours of the 2012 Global Games Jam by Tomasz Kaczmarek with support from Michael Scott, both students at Brunel University. The game was developed in Multimedia Fusion 2. The author would like to thank Tara Voelker, Ian Hamilton and their colleagues from the IGDA Accessibility SIG and Special Effect for organizing the Game Accessibility Challenge.

## 7. REFERENCES

[1] Atkinson, M., Gucukoglu, S., Machin, C.H.C. and Lawrence, A. 2006. Making the Mainstream Accessible: What's in a Game? *Lecture Notes in Computer Science: Computers Helping People with Special Needs*, 4061, 380-387.

[2] Bierre, K., Ellis, B., Hinn, M. Ludi, S. and Westin, T. 2005. Game Not Over: Accessibility Issues in Video Games. Available from: http://www.igda.org/accessibility/hcii2005 gac.pdf; accessed 21 March 2013.

[3] Cavanaugh, T. 2002. The Need for Assistive Technology in Educational Technology. *AACE Journal*, 10, 1, 27-31.

[4] Cavender, A., Trewin, S. and Hanson, V. 2008. General Writing Guidelines for Technology and People with Disabilities. *ACM SIGACCESS Accessibility and Computing*, 92, 17-22.

[5] DeMarco, T. 1986. Controlling Software Projects: Management, Measurement, and Estimates. NJ, USA: Prentice Hall.

[6] Ellis, B. 2012. Accessible Games from the Global Game Jam. Available from: http://www.gamebase.info/magazine/read/the-global-game-jam_786.html; accessed 21 March 2013.

[7] Ellis, B., Ford-Williams, G., Graham, L., Grammenos, D., Hamilton, I., Lee, E., Manion, J. and Westin, T. Game Accessibility Guidelines. Available from: http://www.gameaccessibilityguidelines.com; accessed 21 March 2013.

[8] Entertainment Software Association. 2012. Essential Facts About the Computer and Video Game Industry. Available from: http://www.theesa.com/facts/pdfs/ESA_EF_2012.pdf; accessed 21 March 2013.

[9] Kaczmarek, T., Scott, M., Brunner, J. and Anderson, M. 2012. Mini Free Runner. Available from: http://archive.globalgamejam.org/2012/mini-freerunner; accessed 21 March 2013.

[10] Nielsen, J. and Molich, R. 1990. Heuristic Evaluation of User Interfaces. In *Proceedings of the ACM SIGCHI Conference on Human Factors in Computing Systems* (Seattle, WA, USA, April 1 - 5, 1990). SIGCHI'90. ACM, New York, NY, 249-256.

[11] Nielsen, J. 1989. Usability Engineering at a Discount. In *Designing and Using Human-Computer Interfaces and Knowledge Based Systems*, G. Salvendy and M.J. Smith, Eds. Amsterdam: Elsevier Science Publishers, 394-401.

[12] NPD Group. 2007. Playing Video Games Viewed as Family/Group Activity and Stress Reducer. Available from: http://www.npd.com/press/releases/press 071212.html; accessed 4 May 2008.

[13] PRNewswire, 2008. Disabled Gamers Comprise 20% of Casual Video Games Audience. Available from: http://www.prnewswire.com/news-releases/survey-disabled-gamers-comprise-20-of-casual-video-games-audience-57442172.html; accessed 21 March 2013.

[14] Saltsman, A. 2009. CANABALT. Available from: http://www.adamatomic.com/canabalt/; accessed 21 March 2013.

[15] Smith, K. 2011. Innovations in Accessibility: Designing for Digital Outcasts. In *Proceedings of the 58th Annual Conference of the Society for Technical Communication* (Sacramento, CA, USA, May 15 - 18 2011). STC, Fairfax, VA, 117-127.

[16] Squire, K. 2005. Changing the Game: What Happens When Video Games Enter the Classroom? *Innovate: Journal of Online Education*, 1, 6. Available from: http://website.education.wisc.edu/kdsquire/tenure-files/manuscripts/26-innovate.pdf; accessed 21 March 2013.

[17] Stuart, K. 2012. Global Games Jam Tackles the Issue of Accessibility. Available from: http://www.guardian.co.uk/technology/gamesblog/2012/jan/27/global-game-jam-accessibility; accessed 21 March 2013.

[18] United States Census Bureau. 2008. Statistics on Disability. Available from: http://www.census.gov/hhes/www/disability/disability.html; accessed 4 March 2008.

[19] United States Government. 1998. The 1998 Amendment to Section 508 of the Rehabilitation Act. Available from: http://www.access-board.gov/sec508/status.htm; accessed 21 March 2013.

[20] Voelker, T. and Hamilton, I. 2012. Global Game Jam Accessibility Challenge: Information Packet. Available from: http://gasig.files.wordpress.com/2011/02/globalgamejamaccessibilitychallenge1.pdf; accessed 21 March 2013.

[21] White, G.R. Second Life for the Visually Impaired. 2008. Available from: http://blindsecondlife.blogspot.co.uk/2008_09_01_archive.html; accessed 21 March 2013.

[22] White, G.R., Fitzpatrick, G., McAllister, G. 2008. Toward Accessible 3D Virtual Environments for the Blind and Visually Impaired. In *Proceedings of the 3rd International Conference on Digital Interactive Media in Entertainment and Arts* (Athens, Greece, Sept. 10 - 12, 2008). DIMEA'08. ACM, New York, NY, 131-141.

[23] Yuan, B. Folmer, E. Harris, F.C. 2011. Game Accessibility: A Survey. *Universal Access in the Information Society*, 10, 1, 81-100.